# COLLABORATIVE KNOWLEDGE NETWORKS EMERGENCE FOR INNOVATION: FACTORS OF SUCCESS ANALYSIS AND COMPARISON

Authors

**Nicolas PERRY[1*], Alexandre CANDLOT[2], Corne SCHUTTE[3]**

1 LGM²B, Univ. of Bordeaux 1, Bordeaux, France
2 MNM Consulting, Paris, France
3 GCC, Univ. of Stellenbosch, Stellenbosch, South Africa

## Abstract

RÉSUMÉ. *Les nouvelles technologies de l'information et de la communication favorisent l'émergence de réseaux virtuels centrés sur le partage et l'échange d'expertise. Cet enrichissement croisé des connaissances est un des facteurs clés du développement de nouveaux produits innovants. Or les organisations ont du mal à structurer, piloter, favoriser ou tout simplement reconnaître ces réseaux. Cet article compare quatre niveaux de réseaux collaboratifs, prenant en compte différentes spécificités : type et taille de réseau, contexte culturel et linguistique ou objectifs. Cette analyse fait ressortir des éléments clés pour la réussite et l'efficacité du réseau virtuel, relatif à sa construction et son usage. Il s'interroge aussi sur les facteurs limitant ou néfaste de ces structurations virtuelles.*

ABSTRACT. *New product development needs new engineering approaches. Knowledge is a key resource that impacts traditional, organisational, economic and innovative models. Through NICT (New Information and Communication Technologies), globalisation encourages the emergence of networks that overcome traditional organisation boundaries. International enterprises, European-Community Networks of Excellence or Clusters (competitiveness poles) indicate the need to define a new way of thinking. This new way moves towards an agile, continuous innovative use of knowledge. Based on an epistemic study of knowledge management best practices, four examples show the barriers that can be encountered today. This paper aims defining the key elements that enhance collaborative networks. The analysis of best practices from collaborative environments enables the design of high standard information systems and initiate knowledge ecosystems. A balance between formalism required to share knowledge and fuzziness of social networks triggers new initiatives. This ensures the validity of information exchange through virtual collaboration. It helps to maintain group coherence despite exceeding the natural maximum number of collaborators. Finally the main success or failure factors are highlights and commented to ease the transition from economic-driven to expertise-driven models is then facilitated*

MOTS-CLÉS : *Réseaux collaboratifs, réseaux virtuels, plateforme collaborative, innovation, échange de connaissances*

## 1. Introduction

At the end of the last century, geographical, linear and organisational boundaries moved. The increasing economic network is undergoing the influence of a new resource: knowledge-ware. In the framework of the product development, knowledge networks and changing teams give opportunities to improve the working methodologies and knowledge exchanges between actors. Designers no longer work in a single team, but are involved in several projects with different organisations and partners.

New technologies in information and communication have initiated industrial revolution that needs to redefine the collaborative behaviours [Gardoni, 1999]. Today innovation needs to focus on attractive product/process or services, created through the rational use of knowledge, to keep companies competitive. The empowerment is both technical and organisational on every phase of the Product Life Cycle [Mevellec, 2006].

We are addressing the problem of knowledge networks efficiency measurement and improvement. The efficiency value is based on the information and knowledge exchanges between the designers (i.e. network partners), but also on the relevance and the wealth and benefits for the partners. Different levels are addressed, on the one side the collaborative tools and methods to manage and share knowledge, and on the other side, the share and reuse of these knowledge or information.

Du Plessy highlights the major role of knowledge management value and identifies its drivers [Du Plessy, 2005]. New knowledge browsing mechanisms tend to manage the increasing volume of information, and ease the information consolidation for decision making. But there is still a high risk of loss in knowledge exchange between users due to their own representation mechanisms. A network organisation should limit this risk.

The drift from an economic model to a knowledge model has resulted in two consequences that companies are trying to adapt their policies. Globalisation requires more information for decisions and makes the system hard to optimise. Moreover, it is difficulty to evaluate knowledge and the knowledge exchange value. It points out the limits of the present economic models and the need for the integration of new architecture for knowledge management systems and business process management systems [Jung et al., 2007].

A way to face these problems is foreseen in collaborative evolution, leading to symbiotic networks, breaking old habits and initiating new ways of sharing knowledge. Ontologies facilitate the integration of knowledge and expertise by structuring and formalising the exchanged information in the IT environments [Du Preez et al., 2004, Candlot et al., 2005]. It aims answering the saturation of information and refocus on the contents instead of the flows.

For organisations, external knowledge becomes more important than internal knowledge. For example customer relationship and new products definition has forced organisations to strengthen their relationships [Du Plessy and Boon, 2004]. But, organisations are neither ready for these new interactions nor favour or reward actor involvement. The tendency seems to drift from a value centred on artefacts to a value centred on the flow of artefacts [Fogel, 2005].

For the designer point of view the benefits come from the mix and exchanges of knowledge from different domains. Indeed, working on new products development or searching for innovation, the information speed up the validation of concepts or the creativity on solutions. From this perspective, knowledge-based projects are analysed in order to understand collaborative work. In this paper we are studying collaborative interactions in five real-case study: a research team, a long distance bilateral research collaboration, a French project financed by the Ministry of Industry and two European Networks of Excellence. The expertise partners' background, the size of networks, the collaborative processes and tools are indicators to the understanding of the benefits and disadvantages of these projects [Petiot et. al., 2001].

This paper will first present the evolution from customer-oriented enterprises to knowledge-oriented networks. Therefore communities emerging from the use and the share of new tools derived from NICT are the pioneers of a new knowledge-based economy. New value models create another perception of information validity. Symbiotic network best practices result from these models.

## 2. Semantic positioning

Computerisation transforms work environments. Information used to be inseparable from its material support (usually the paper). The new environment breaks these borders. The huge amount of available information raises the question of what is essential and relevant. A search for epistemological justification has begun. [Zuniga, 2001].

First of all, what is a document according to new technologies ? The document is no longer a static object. IT revolution separated the capitalisation function to the knowledge sharing function. Internet evolutions and mark-up technologies separate structures from contents [Bachimont & Crozat, 2004]. Consequently, it allows multiple reconstructions depending on the user's perception using the editor's tools possibilities and the multi-layer technologies. A paradox appears between the previous functions to maintain leadership and its openness to guarantee the exchangeability with other computers [Fogel, 2005]. In addition annotation tools, based on "Web 2.0", put the user no longer just a customer but as an actor that assesses and gives feedback on content [Frank and Gardoni, 2005] [Keraron et. al., 2006]. Finally, knowledge is created through the interaction between computers and users, depending on the type of media explored. It moves from the understanding of concepts contained in a text to the scanning of multiple-content pages [Fogel, 2005]. By an interpretation function and a process of reconstruction, the data contained in the media are activated to give back the meaning [Searle, 1980].

But it remains the human representation and interpretation of the documents. Several research communities address the problem of knowledge representation through models, including knowledge managers and business activity modellers. By comparing English and German translations of the word "representation", we can classify their work. Figure 1 illustrates Kant's and Frege's discussions and proposes four entries for "representation" [Cassin, 2004]. It denotes a reconstruction of meaning from data to knowledge through perceived information [Bernard et. al., 2005]. Table 1 presents a consolidated view of different modelling approaches (KM, BPM, GERAM) based on their similarities to German epistemology. A simplified vision of cognitive processes emerges. Knowledge has to be de-structured to be shared. All methodologies presented in Table 1 propose, on the one hand, to ease the enrichment cycles and, on the other hand, to limit the degradation of knowledge between users [Candlot 2006].
As a consequence we cannot keep a fixed reference. The collaborative platform should continuously adapt to an actor's perception, understanding and needs. The de-structuring/structuring exchange process means that the value is in not in knowledge but in knowledge exchange. Ontology-based tools help map and stimulate potential interactions between actors.

## 3. Knowledge validity for experts exchanges

This balance between information structuring and use flexibility founded partial solutions among which for instance indexes, summary, keywords or tables of content. For a desynchronised and now numeric transfer of expertise, the degradation of knowledge in data necessitates new navigation tools to correct the lack of context for interpretation. The multi-user approach of collaborative platforms or networks requires a common language between experts, to confirm relevance, authority and confidence in resources and the information therein. These terms can be defined as follows:
- Validity = Relevance + Authority + Confidence

- Relevance = corresponds to my interest
- Authority = has been assessed by a mediator that I am confident in; is recognised by a large community; could be assumed as proof
- Confidence = seems interesting to me; is something I personally trust

These concepts should help users to assess in real time the validity of the observed knowledge network. The use of these terms appears progressively in different tools. Similar language-synchronisation and document-navigation tools illustrate the evolution of indexing tools towards a naturally valid and dynamic system. It goes from the terminology (list of terms), the glossary (list of definitions), the taxonomy (structured list of definitions) the thesaurus (semantic and structured groups of definitions organised in networks) and the ontology (objective networks of defined concepts).

Using ontologies, engineering reaches an inter-subjectivity that becomes the local objective of a community [Zuniga, 2001]. These agreements enable multi experts to reach consensus and smooth misunderstandings and concept gaps. As a result, three main research categories arise. First the research of consensual vision between different stakeholders. Definitions are slightly different from one expert to another. The small gap is often enough to stop convergence [Guarino, 1995]. A second research field focus on the model comparison [@Metis] [@INTEROP]. Ontology is a way to align the models. The third field of research deals with the artificial intelligence and the decision-making or case-based reasoning. Ontology is used as an indexing tag library at a high semantic level. But it remains the difficulty of the common analysis reference definition and the construction of the initial common understanding.

Most attempts at using ontology finish in a cul-de-sac due to an imprecise understanding and definition. The tool deployment becomes an infinite fruitless task, due to the confusion in goals definition between objectivity and inter-subjectivity. It appears necessary to create a tool and performance measurements that helps users to quickly assess the relevance of information, through context, interpretation and meaning. Actually the most used search engines on the web focus on time-to-information and most-visited places despite relevancy. Their practice transforms the relevance into worldwide confidence. Validity shifts towards confidence. The most valid information is evaluated less on its relevance than by the number of people who share it.

It results that the knowledge exchanges with little context definition (document definition issue, representation mechanism) lead to misunderstanding and Knowledge loss. Thus, it is time to propose tools to reduce the semantic reconstruction gap between virtual representations and real systems (validity – sum of relevance, confidence, authority – and ontology). Design collaboration implies multiple points of view. The value of new knowledge comes from their synergy and not from their reduction to a single common view. The task of structured tools (based on numerical technologies) is to absorb and redirect potential knowledge and ease its absorption for new users. But the selection criterion is often not the relevancy but the frequency. Flows become more strategic than contents. The new tools will contribute to establishing a new authority that could be the base of collaborative ecosystems [@CORDIS].

## 4. Ecosystems and Knowledge Networks

Knowledge Networks are defined by Du Preez and al. [Du Preez 2008] as:
*"A Knowledge Network signifies a number of people and resources, and the relationships between them, that are able to capture, transfer and create knowledge for the purpose of creating value. An Integrated Knowledge Network spans all domains,*

*communities, and trust relationships with the goal of fostering sustainable innovation that will continue to promote the competitiveness of its users.*"

These Kwoledge Networks environment are built on virtual teams (and the inhibitors for collaboration), innovation (and its (inter)dependence on knowledge), the knowledge creation process, innovation project methodologies, collaboration and evaluation of the collaboration improvement using the knowledge networking. However, each domain individually and collectively impacts on the success of innovation projects. A common framework that integrates these concepts into a single methodology will be useful.

The capitalisation of some kinds of knowledge has already evolved through several generations of management. Companies have first focused on the product. Secondly, the process and project management of innovation becomes the key factor of performance. Thirdly the whole company has been considered as an innovative place, and not only a production area. In the fourth evolution, innovation overcame the boundaries to reach the client, to take its opinion into account and to acquire an agility to anticipate its choices. This evolution described in [Amidon, 1997] [Savage, 1991] ends by a fifth generation that corresponds to global, innovative and symbiotic networks of knowledge workers. These steps are summed up in Table 2.

The corresponding evolution of tools moved from the Personal Best Practices, to Shared Databases, Expert tools (AI, KBS, Road maps, Master Plan…), NICT, Groupware (Internet, Networks, ERP), then Ecosystems (Collaborative Platforms, Community Networks) [Gunnlaugsdottir, 2003] [Benbya et al., 2004] [Rodrigez and AlAshaab, 2005] [Nia et al., 2006].

We have highlighted the evolution of knowledge-sharing in companies from a human-centred use to a decentralised network. NICT evolutions coupled with ontology-based approaches are solutions to help breakthrough innovations. The following sections will analyse new knowledge-based networks which are experiments for new collaboration environments and provide some examples of best practises.

## 5. Examples

The next section will present and analyse some different designer network's configurations. Table 3 introduces the five networks, from the single team to the European networks of excellence. They are compared each other in order to highlight the impact on collaborations. The symbiotic network relies on mutual recognition of partners and user-friendly tools in order to work together. Moreover, information validity needs relevance of response, confidence in the collaborator and mutual authority recognition in the network. The analysis of the different networks is based on several parameters: the number of collaborators involved and their space distribution, the context with the expertise background and their differences, and the strategy definition of the partners and the potential agreement on common ones. We will try to highlight the good practices and we will discuss some results regarding the efficiency of the network.

*5.1. Team level*

The teams are the smallest cells of collaboration and are composed of geographically and thematically close workers. Different collaborators (here in the case of a research group) working on their own specific subjects enrich a common general research domain.

- Formalisation: day-to-day contacts and informal meetings form the core group. Discussions seem to be enough to maintain a common concept-sharing and formalisation. The informal exchanges stimulate information flows between experts and favour serendipity. However, for a specific domain, when more in-depth work is carried out (technically or conceptually), it seems that a formal synchronisation phase is required to upgrade the synergy of all the collaborators.
- Objectives and strategy: frequently poor definition of objectives for the global team leads to limited collaboration within the team on elementary tasks.
- Collaborative platform: it is difficult to involve people in the use of common tools. Workers argues on PC versus Mac hardware, Microsoft© versus Open Source editors, different software or even different updates. Imposing software ignores the specific skills of each worker and may address a low reference. The drawbacks are the price of software and files sharing, problems that are addressed by all Internet businesses. Emerging solutions have still to be assessed and partners' wishes taken into account. Heterogeneity and diversity is a wealth and should be preserved.

At this level, the validity of information shared is automatic and every question is immediately solved by proximity. It favours serendipity. Human flexibility enables the creation of synergy even with a poor definition of objectives. The community of knowledge is maintained on a day-to-day basis and is enforced or adjusted by the physical presence of the collaborators.

*5.2. A long-distance research team collaboration*

The analysis of the collaboration between two French and South African teams [@IVGI] [@GCC] highlights the efficiency of a first round of physical meetings and their fruitful and relevant results.
- Formalisation: a first round of mutual description and understanding was necessary. This step started with the writing of a green paper describing both sides' concepts and the different points of view and areas of interest. This was the result of two man/months (equivalent) physical meetings. It creates a common ontology from the conceptualisation of each culture.
- Objectives and strategy: no specific target are expected, so the network is only reactivated for specific actions, such as student internships, co-written papers, and collaborative tools specification and tests done before and during the exchange. Long-term actions are harder to do, as day-to-day business backlog is time consuming.
- Collaborative platform: due to cultural diversity, physical exchanges help to understand, share and agree on expertise or viewpoints of the same issue. A web based collaborative environment for document management has been set up for distant working collaboration (Webeden™). Its use is sporadic as the collaboration only works when partners are in the same place.

Information validity took one year of collaboration to develop knowledge on the partners' expertise. The relevance and confidence of shared documents and the emerging of cross-combined knowledge has led to mutual authority recognition at both sides. The network has potential, but it is only activated by immediate requirements that lead to action. Consequently, methods should be found to prepare the networks for the action. Ecosystems should be defined in this perspective.

*5.3. National research / industrial projects*

The French national USIQUIK project [@USI] faces more organisational problems:
- Formalisation: the wish is to to model the knowledge used in the projects. A first UML based proposal (activity, sequence and class diagrams) was not flexible enough to follow partner's evolution. The second enriched MOKA ICARE files [Moka, 2001] with organisation data [Ammar, 2005]. In both cases, the dynamic change was not manageable if the changes were not directly made in workers' environments.
- Objectives and strategy: each partner maintains a high degree of freedom. The project was difficult to manage due to a lack of hierarchy. Moreover, each partner's responsibilities are fuzzy or are rejected or ignored. As a result, clear common working methodologies are not used.
- Collaborative platform: despite the formalisation of the complete project's concepts and phases (manufacturing terms and project steps using UML formalism) and the set up of a web site and forum, each partner works with the minimum of interaction with the others. Partners spontaneously recombine their relation in pairs and seldom share their visions and work.

If the symbiosis between network members cannot be ensured by a common reference, this network cannot benefit from its potential. The relevance of the common reference should be ensured directly by all the project stakeholders. The lack of recognised authority and confidence of initially shared information has broken the trust between partners and induced a divergence of objectives. The reference model should maintain a consensus agreement on objective evolutions.

*5.4. European Network of Excellence*

Two networks of excellence from the FP6 of the European Community constitute an analysis panel for bigger networks [@VRL-KCiP] [@INTEROP]. In these cases, the number of partners creates a new difficulty to face. In both cases, the large size causes smaller sub-groups to emerge. We recognise that different aspects of information validity are pre-requisites for the mutual recognition of actors

On the one hand, in VRL-KCiP, the partners share the production domain of expertise of the CIRP community. Subgroups result either from already existing networks (from previous European experiences for example) or from existing structures (CIRP structures for example) or previous collaborations. These subgroups are based on already mutually recognised confidence and authority.

In INTEROP there is the combined expertise of enterprise modelling, ontologies and software architecture and platforms. As a result, subgroups spontaneously resulted based on similar expertise and partners that already worked together. In this case, the groups are formed based on relevance and confidence.

In these two different ceses, the small groups progressively reconstruct an environment which is propitious to information validity. A big challenge is to regularly break and recombine the groups in order to encourage cross knowledge fertilisation and make the global network efficient.
- Formalisation: global mutual understanding is needed for knowledge sharing, but this task is time consuming. The INTEROP knowledge map and glossary are steel being built after almost three years. Considering this amount of work, the attempt may be difficult. More than 2000-shared terms in the glossary are almost unmanageable. VRL-KCiP started a similar task. Here is the question of balance between scientific exhaustiveness (ontology) and engineering efficiency

(conceptualisation) of the knowledge reference. Optimisation needs to be done regarding the size of the studied domain, the number of partners and the objectives to fulfil. These three parameters are interlinked. The increase of partner numbers will not directly increase the potential of the network for each partner, an optimum appears. The limit depends on the number of confident people from which sustainable interaction could be expected. Confidence is the only prerequisite of validity shared by the two networks. Thus, these emerging ecosystems imply a balanced distribution of influence as a key to objective fulfilment.

- Objectives and strategy: Considering working relations, no indicators are available to measure the efficiency of the collaboration, except the final deliverable agreement. Thus tasks and work-group management only rely on partner involvement and goodwill. At the global management level it is not possible to follow all the actions. Thus and because of the size, some works could be redundant and even sometime useless. Moreover, partners' involvement depends on benefit feedback. The first phase is critical. It should be aimed at building a win-win environment. So the network reinforces the links for partners who are already collaborating. In the middle term, the network benefits new partners that first have to weave connections and be recognised as valid by others.
- Collaborative tools: one of the advantages of working in such a big European project is to understand Western European cultural habits, learn to decode and work with them. Germanic rigor, Anglo-Saxon pragmatism, Latin adaptation, Scandinavian synthesis, are pluses and minuses that must be combined to become a strength and not a cultural wall. Understanding this nuance between partners helps knowledge-sharing and collaborators confidence in such a way that answers can be customised depending on the country in question.

VRL-KCiP chooses a collaborative tool without the full agreement of the partners and struggles to make people use it. Other solutions are sometimes used in parallel. Moreover, one unique partner owns the database and access to configure the environment. Due to the lack of confidence between partners, the legitimacy is discussed and decisions do not reach global agreement and involvement.

INTEROP has developed a web-portal for document repository and information sharing. Its use is easy even if navigation is difficult because of the project size. No partner personally owns the database, the service is rented to a company.

So the following recommendations should be followed: neutral database localisation, full web interface efficient enough to avoid duplication on personal computers and to facilitate browsing in this huge knowledge space.

The Interop network highlights partners' authority recognition and the need of pragmatism instead of exhaustivity. On the other hand, VRL-KCiP network emphasises the need of group redeployment and the slow but inexorable common understanding emergence. Both show the limits of numerous groups and the difficulty to efficiently share with all partners. A controlled size of these knowledge networks should enable an optimum configuration to be reached.

**6. Discussion**

The major assets of a company move from financial to human. The improvement in knowledge capital has triggered the creation of many knowledge management projects in companies. The next step, for enterprise capitalisation awareness, is to give value and take the benefits of the network. The collaboration of experts, most of the time coming from different structures, embodies more knowledge than the sum of each expert individually. It feeds the innovation process.

Consequently, practices in organisations have to change to integrate the new informal groups which are setting up around similar objectives. First, teams or project leaders have to be mature enough in order to identify the most efficient connection to develop. To favour this team empowerment, the structures have to decentralise information spread, technically and hierarchically support it, and promote the networking gains.

Documents, that are the keystone of capitalising, sharing and spreading, must be analysed and profiled for an efficient knowledge enrichment (to ease the innovation process) and degradation (to limit the interaction losses). Whatever the mutual efforts defining collaborative platforms, cultural gaps will still remain that cannot be deleted. This difference should be taken into account to favour the set up and to take the best of the sources from wherever they come.

The five examples were analysed on different aspects: number of working partners, their domain expertise and their background. The result of this analysis highlights the need of validity i.e. the sum of confidence between partners, mutual authority recognition and relevancy of the information shared. The analysis of knowledge formalisation, network strategy and collaborative platforms used, points out some best practices (see Table 4). The latter ease the propagation of information validity in virtual networks which are larger than naturally efficient group sizes.

First of all, knowledge formalisation reveals interaction areas. The experiences highlight the importance of physical meetings with face-to-face discussions. Human beings need to synchronise their views, share methods and tools before enlarging their circle of confidence. Secondly, it is essential to reach mutual understanding regarding agreement on terms and concepts before reaching the information validity level. Based on our experience, regardless of the number of partners, their expertise and background, an exhaustive glossary, taxonomy or even ontology is difficult to achieve (despite difficulties to define this tool). These tools will be efficient only if they are closely aligned with the network objectives. Users should have direct feedback on their time invested. The sum of all these actions will, in the long term, favour the dissemination of valid information.

A second category of best practices deals with network objectives and strategy. Each involved expert has a personal strategic orientation. It is becoming more and more important to explain and integrate the alignment of all these objectives within the virtual group. Personal or network strategy may form boundaries that are not objective with administrative ones. These differences risk alienating some partners and then killing the symbiosis. Again, a fair participation should ensure feedback to actors. The collaborative platform should help to solve this main issue.

Some easy requirements could ensure a good start to the network collaborative platform. Among them, a neutral hosting guarantees, an equal distribution of responsibilities and the respect for intellectual property. A full internet based web solution keeps interaction potentially active in order to maximise fruitful opportunities. Ergonomics of its interface and browsing facilities will ensure an instinctive use of assessing relevance, confidence and authority of a problematic analysis. This validity guarantee increases the synergies.

## 7. Conclusions and perspectives

Common interest collaborative networks are new opportunities to benefit from the NTIC and knowledge exchanges within and outside the structures.

In the context of globalisation, the optimisation of expertise potential relies on the development of sustainable synergies. It implies an in-depth redefinition of working interactions. The NICT has given an undeniable value to these groups. Organisations have to adjust their management, their information privacy policies and their innovation processes. By replacing previous financially-based consortia, where relations were less dependent on core competencies, these knowledge-based ecosystems have become a major centre of interest for future extended companies.

**Biography**
A.CANDLOT

Alexandre Candlot is a research engineer focusing on Process Planning expertise, and thus on knowledge management. Working in the Research Institute of Communication and Cybernetics of Nantes, he has defended a PhD based on the analysis of industrial case studies and has identified ways to facilitate expertise integration within numerical or virtual environments. He is now Senior Consultant in MNM Consulting, dealing with organisational and knowledge based projects and IT specification.

N.PERRY

Nicolas Perry received his PhD in Mechanical Engineering from the University of Nantes and Ecole Centrale de Nantes (Fr) in 2000. He is an Associate Professor at Ecole Centrale de Nantes and works in the Research Institute of Communication and Cybernetics of Nantes. His research topics focus on Virtual Engineering, Knowledge Management and KBE as Decision Tools for Engineers applied to Cost Management.

C. SCHUTTE

Corne Schutte is currently a senior lecturer and PhD student at the Department of Industrial Engineering, University of Stellenbosch in South Africa. He works in the Enterprise Engineering research group and his research topics focus on the setup and use of integrated knowledge networks with the purpose of enhancing innovation.

**Figures**

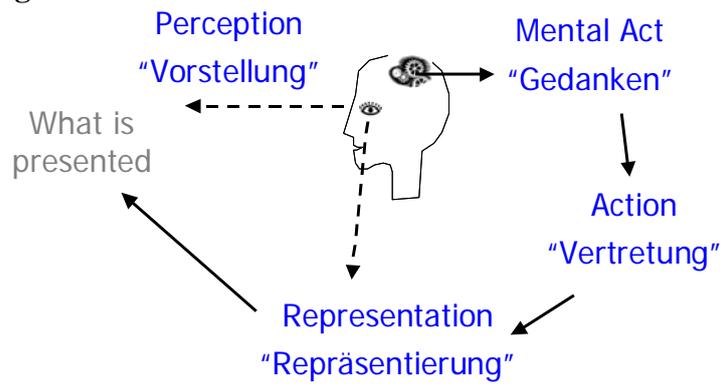

Figure 1 : The four German translations of representation

**Tables**

| | | ENRICHMENT | | DEGRADATION | | | |
|---|---|---|---|---|---|---|---|
| Philosophy | Kant / Frege | Vorstellung | Gedanken | Vertretung | Repräsentierung | | German Epistemics |
| | Approximative Translation | Sensation | Thought | Proxy | Representation | | English Epistemics |
| Consolidated KM Meanings for "Representation" | | Perception | Mental act | Action | Vehicle | | Point of view |
| KM (Knowledge Management) | Nonaka SECI Model | Internalization | Combination | Externalization | Socialization | | Social aspect |
| | Ermine Marguerite Model | Learning | Creation | Capitalisation | Sharing | | Continuous improvment |
| | Grundstein GAMETH Model | Identify | Actualise | Preserve | Valorise | | KLC maintenance |
| BPM (Business Process Modelling) & GERAM (Generalised Enterprise Reference Architecture Model) | PERA | Identify | Analyse | Build | Operate + Kill | | Life Cycle |
| | CIMOSA BPM 4 Objectives | acquire explicit knowledge about the business processes | support the decision making activities | exploit this knowledge in BPR projects | ease interoperability | | Standardisation / interoperability |
| | GRAI Integrated Method | Identify | Model | Specify | Implement | | Decision Flow |
| | PMI (Muench) Representative Software Development | Identify | Design | Construct | Evaluate | | Project Management |
| KBE monitoring | Candlot Proposal | Understand | Enrich | Re-use | Explain | | Change management |

(Column separators labeled vertically: DATA | INFORMATION | KNOWLEDGE | INFORMATION | DATA)

Table 1: Simplified vision of different model-management methodologies consolidated on German epistemology for "Representation"

| Enterprise Generations [Amidon] | Capability Maturity Model | Synthesis | Project Decision-Making Dimension |
|---|---|---|---|
| | Level 1 « initial » | "Fire Fighting" | |
| | | CRAFT | |
| 1st - Asset: Produit (unpredictable change) | Level 2 « reproductible » | "hand made" | |
| | | INFORMATICS: Human - Computer | Operational |
| 2nd - Asset: Project (Interdependance) | Level 3 « define » | Procedures, PC, Software Tools | |
| | | INFORMATICS: Human - Computer - Human | Tactical |
| 3rd - Asset: Enterprise (Technology & Systematic Management) | Level 4 « master » | Interoperability, local network, PLC integration | |
| | | INFORMATICS: Human - Computer - Groupe | Strategic |
| 4th - Asset: Client (global change, experience feedback) | Level 5 « optimise » | Behaviour, mystake anticipation | |
| | | INFORMATICS: "Knowledge Workers" | |
| 5th - Asset: Knowledge (participative innovation, symbiotic network) | | autonomous learning, global network, synergy throughout fences (country, enterprise…) | |

Table 2: Evolution of knowledge management maturity

|  |  | TEAM | South African collaboration | French National Project USIQUIK | NOE VRL KCiP 6th FP | NOE INTEROP 6th FP |
|---|---|---|---|---|---|---|
|  | Approx nb of partners / people involved | 1 Team / 10 People | 2 Teams / 20 People | 8 Teams (5 lab, 2 indus, 1 gov) / 30 People | 30 Partner Teams / 200 People | 50 Partner Teams / 350 People |
|  | Background | Close profiles | Geographical en cultural gaps | Res Lab Industrial | Research Lab Emerging European culture | Research Lab Emerging European culture Connection with other networks |
|  | Expertise | Close | Industrial Engineering Software development | Manufacturing Software development | Production Knowledge management | Enterprise modeling Ontologies Software architecture and platform |
|  | Objectives | Industrial Engineering for Virtual Engineering | Encourage research exchange between the two teams | Automates Process Planning | Virtual Research Lab. | System Interoperability |
|  | Clarity of Objective Definition | Learning by doing | Free to evolve | RNTL Project, difficult to understand | Clear | Prospective research difficult to figure out |
| IMPACT ON THE COLLABORATION | Size Impact | No | No | No | Yes, smaller groups emerged between people used to work together | Yes, smaller groups emerge, according to expertises |
|  | Working Relations | Single person task | Limited to physical meetings | Difficulty to establish responsibilities | Even if the network works, Struggle to define middle term objectives | Tasks difficult to drive, unclear indicators, useless or redundant jobs |
|  | Collaborative Platform | No | Yes, Webeden (not very used) | Web site & Forum (both not used) | SmartTeam / Webeden (no clear and agreed choices) | Web site (used, but hard to navigate) & Forum (not used) |
|  | Cultural Impact - Country | No | High Distance (rare meetings) | No | Benefiting from CIRP influence, attracting world-wide partners | Concentrated on Occidental European, a culture emerge from various customs |
|  | Cultural Impact - Expertise | No | Positive despite different expertises | Deep differences in problem solving | Same concepts with different words and visions, difficulty to reach consensus | Strength emerges after a first barrier of the diversity |
|  | Ontology | Informal Synchronisation | Green Paper resuming mutual concepts | UML Diagrams (hard up-dating, not used) | Knowledge Map & Taxonomy (Hard to promote among partners) | Knowledge Map & Glossary (Huge, unachievable, no use identified) |

Table 3: Comparative benefits and gaps for different networks of experts

|  |  | TEAM | South African collaboration | French National Project USIQUIK | NOE VRL KCiP 6th FP | NOE INTEROP 6th FP |
|---|---|---|---|---|---|---|
| Value | First Best Practice | Proximity | Virtual Proximity | Objective evolution adaptation | Partener autority recognition | Task force redeployement |
| | Second Best Practice | Flexibility | Mutual understanding | Represantation constant sharing | Progmatism instead of exhaustivity | Slow but inexorale common understanding emergence |

Table 4: Two main best practices learned from each case